\documentclass[conference]{IEEEtran}
\IEEEoverridecommandlockouts
\usepackage{cite}
\usepackage{amsmath,amssymb,amsfonts}
\usepackage{algorithmic}
\usepackage{textcomp}
\usepackage{xcolor}
\usepackage{comment}
\usepackage{graphicx,pstricks}
\usepackage{graphics}
\graphicspath{{img/}}
\usepackage{booktabs}
\usepackage{tabularx}
\usepackage{caption}
\usepackage{array}
\usepackage{cancel}
\usepackage{multirow}
\usepackage[most]{tcolorbox}

\tcbset{
enhanced,
breakable,
left=2mm,
right=2mm,
attach boxed title to top left={
xshift=0.2cm,
yshift=-5mm,
yshifttext=-1mm
},
top=4mm,
colback=orange!5!white,
colframe=orange!100!black,
colbacktitle=orange!100!black,
boxed title style={size=small,colframe=orange!100!black}
}
\usepackage{setspace}
\usepackage{hyperref}
\newenvironment{mybox}[1]{%
\begin{tcolorbox}[title={#1}]%
\setstretch{0.95}}{
\end{tcolorbox}
}
\usepackage{todonotes}

\def\BibTeX{{\rm B\kern-.05em{\sc i\kern-.025em b}\kern-.08em
    T\kern-.1667em\lower.7ex\hbox{E}\kern-.125emX}}

\usepackage[all=normal,floats,mathspacing,wordspacing,charwidths,indent,lists,bibnotes]{savetrees}

\begin{document}

\title{Enhancing AI-based Generation of Software Exploits with Contextual Information}
%{\normalsize \textbf{(Practical Experience Report)}}

%\author{Anonymous Author(s)}

\begin{comment}

\author{\IEEEauthorblockN{1\textsuperscript{st} Given Name Surname}
\IEEEauthorblockA{\textit{dept. name of organization (of Aff.)} \\
\textit{name of organization (of Aff.)}\\
City, Country \\
email address or ORCID}
\and
\IEEEauthorblockN{2\textsuperscript{nd} Given Name Surname}
\IEEEauthorblockA{\textit{dept. name of organization (of Aff.)} \\
\textit{name of organization (of Aff.)}\\
City, Country \\
email address or ORCID}
\and
\IEEEauthorblockN{3\textsuperscript{rd} Given Name Surname}
\IEEEauthorblockA{\textit{dept. name of organization (of Aff.)} \\
\textit{name of organization (of Aff.)}\\
City, Country \\
email address or ORCID}
\and
\IEEEauthorblockN{4\textsuperscript{th} Given Name Surname}
\IEEEauthorblockA{\textit{dept. name of organization (of Aff.)} \\
\textit{name of organization (of Aff.)}\\
City, Country \\
email address or ORCID}
\and
\IEEEauthorblockN{5\textsuperscript{th} Given Name Surname}
\IEEEauthorblockA{\textit{dept. name of organization (of Aff.)} \\
\textit{name of organization (of Aff.)}\\
City, Country \\
email address or ORCID}
\and
\IEEEauthorblockN{6\textsuperscript{th} Given Name Surname}
\IEEEauthorblockA{\textit{dept. name of organization (of Aff.)} \\
\textit{name of organization (of Aff.)}\\
City, Country \\
email address or ORCID}
}
\end{comment}

\author{
    \IEEEauthorblockN{Pietro Liguori\IEEEauthorrefmark{1}, Cristina Improta\IEEEauthorrefmark{1}, Roberto Natella\IEEEauthorrefmark{1}, Bojan Cukic\IEEEauthorrefmark{2} and Domenico Cotroneo\IEEEauthorrefmark{1}}
    \IEEEauthorblockA{\IEEEauthorrefmark{1}University of Naples Federico II, Naples, Italy
    \\\{pietro.liguori, cristina.improta, roberto.natella, cotroneo\}@unina.it}
    \IEEEauthorblockA{\IEEEauthorrefmark{2}University of North Carolina at Charlotte, Charlotte, NC
    \\bcukic@charlotte.edu}
}

\maketitle
\thispagestyle{plain}
\pagestyle{plain}

\begin{abstract}
This practical experience report explores Neural Machine Translation (NMT) models' capability to generate offensive security code from natural language (NL) descriptions, highlighting the significance of contextual understanding and its impact on model performance. Our study employs a dataset comprising real shellcodes to evaluate the models across various scenarios, including missing information, necessary context, and unnecessary context. The experiments are designed to assess the models' resilience against incomplete descriptions, their proficiency in leveraging context for enhanced accuracy, and their ability to discern irrelevant information.
The findings reveal that the introduction of contextual data significantly improves performance. However, the benefits of additional context diminish beyond a certain point, indicating an optimal level of contextual information for model training. Moreover, the models demonstrate an ability to filter out unnecessary context, maintaining high levels of accuracy in the generation of offensive security code.
This study paves the way for future research on optimizing context use in AI-driven code generation, particularly for applications requiring a high degree of technical precision such as the generation of offensive code.
\end{abstract}


\section{Introduction}
\label{sec:introduction}

In recent years, the integration of AI-code generators into the programming workflow has marked a paradigm shift, significantly streamlining coding tasks and facilitating the interpretation of complex instructions through natural language (NL) using advanced ML models. This evolution has been particularly transformative in fields that demand high technical skills, such as \textit{offensive security}, where the automation of coding tasks promises to boost productivity and innovation~\cite{mirsky2022threat}.

Offensive security, with its emphasis on developing \textit{proof-of-concept} attacks to uncover and exploit vulnerabilities, occupies a critical junction in cybersecurity efforts. It aids in understanding the mechanics of attacks, motivating timely patches and mitigations~\cite{avgerinos2014automatic}. However, the manual crafting of exploit code, given its dependence on deep system-level knowledge, presents a considerable bottleneck. Herein lies the appeal of automatic exploit generation (AEG): an AI-driven solution to enhance the efficiency of security analysts by generating functional exploits for security assessments.

Recent advancements have seen AI-based code generators, employing Neural Machine Translation (NMT) models, make significant strides in translating NL descriptions (\textit{intents}) into precise programming \textit{code snippets}~\cite{tufano2019learning,mastropaolo2021studying}. This capability is particularly noteworthy within the offensive security domain, where code generators translate intents detailing the exploitation of system vulnerabilities into actionable code~\cite{natella2024ai,liguori2024power}.

Yet, this domain's complexity introduces distinct challenges. The need for a stringent vocabulary to articulate low-level operations essential for exploiting system vulnerabilities means that the variability inherent in technical descriptions can severely impact the functionality of generated code. Variations in developers' technical knowledge, terminology use, and description specificity can lead to significant semantic discrepancies~\cite{wang2013study,steidl2013quality}. Such variability underscores the necessity for code generators to exhibit \textit{robustness}, ensuring effective operation across a wide spectrum of sentence structures, vocabularies, and detail levels. The absence of this robustness could limit code generators' real-world applicability, constraining their usbaility in practice~\cite{DBLP:conf/icse/MastropaoloPGCSOB23,improta2023enhancing}. 
Hereby, exploit generation is pivotal in bridging the gap between theoretical NMT capabilities and practical, high-stakes security applications. 

This practical experience report offers a critical examination of the potential and limitations of AI-code generators when faced with NL descriptions of varying accuracy and detail in the realm of applications requiring high levels of precision and contextual awareness, such as in the development of offensive security code. 
More precisely, this work delves into the potential benefits of training the models to comprehend the context of a sentence, aiming to discern the extent to which context understanding can compensate for the ambiguities of the NL. 
%To this aim, we employ concatenation as a means to provide contextual information, a solution that requires no alterations to the fundamental architecture of the models.
To this aim, we employ a prompt-engineering solution that feeds models with context information (\textit{contextual learning}) by using the concatenation of inputs, i.e., by merging previous NL descriptions with the current one~\cite{DBLP:conf/discomt/TiedemannS17,DBLP:conf/emnlp/WangTWL17,scherrer-etal-2019-analysing,voita2018context}. 

By harnessing the power of contextual learning, we demonstrate how AI code generators can significantly improve in translating NL descriptions into code for software exploits. Our study performs an extensive evaluation involving six state-of-the-art NMT models, each assessed for its ability to navigate the complexities of generating code from NL descriptions that vary in detail and context. Furthermore, the construction of a unique dataset from real-world exploits for model training and evaluation underscores the practical relevance and applicability of our findings.

To conduct our study, we identified three research questions (RQs) aimed at exploring the capabilities of NMT models specifically within the domain of offensive security. These questions focus on the models' ability to handle missing information, their ability to leverage contextual learning to enhance code generation, and the impact of unnecessary context on their performance. Central to our analysis is the examination of how these factors influence the accuracy of the generated code, pivotal in offensive security applications. Additionally, we investigate the potential of these models to discern and filter irrelevant information, i.e., a critical capability for maintaining high performance in complex security scenarios. \tablename~\ref{tab:findings} presents a concise summary of our key findings, offering insights into the optimization of NMT models for the requirements of offensive code generation.

In the following, 
Section~\ref{sec:related} discusses related work;
Section~\ref{sec:problem} illustrates the problem statement;
Section~\ref{sec:research_study} describes our research study; 
Section~\ref{sec:setup} details our experimental setup; 
Section~\ref{sec:experimental} presents the experimental evaluation;
%Section~\ref{sec:discussion} discusses our findings;
Section~\ref{sec:threats} discusses threats to validity;
Section~\ref{sec:conclusion} concludes the paper.

\begin{table}[t]
\centering
\caption{Main findings of our research study.}
\label{tab:findings}
\footnotesize
\begin{tabular}{
>{\centering\arraybackslash}m{1.5cm} |
>{\centering\arraybackslash}m{6.5cm}} 
\toprule
\textbf{Analysis} & \textbf{Main Findings} \\ \toprule
\textit{Missing Information} &
\begin{itemize}
  \item Models struggled more with highly technical or specific operational commands when details were absent, underscoring the need for detailed NL descriptions in security-related tasks.
\end{itemize} \\ \midrule
\textit{Contextual Learning} & \begin{itemize}
    \item Introducing additional context significantly improved model performance, with a notable enhancement in generating complex, multi-step security protocols.
    \item There was a diminishing return on performance when extending context beyond a single preceding intent, indicating an optimal level of contextual information for effective learning.
\end{itemize} \\
 \midrule
\textit{Unnecessary Information} &
\begin{itemize}
  \item Unnecessary information did not degrade model performance; models were able to maintain high performance, indicating effective filtering of irrelevant information.
\end{itemize} \\
 \bottomrule
\end{tabular}
\end{table}

\section{Related Work}
\label{sec:related}
% related work su software exploits 
% related work su context aware generation, sia per NL sia per code comment gen
% related work on SOTA language models

%\imp{sarebbe interessante fare uno studio encoder-decoder VS decoder-only con contexual learning. abbiamo visto che E-D sono migliori, ma si può notare anche che modelli con architetture diverse hanno risultati diversi in base all'exp. e.g., E-D vanno meglio con contesto utile, mentre D-O vanno meglio con contesto inutile. posso usare un altro E-D (AST-T5, PLBart) e un altro D-O (CodeParrot, PyCodeGPT). }

%\cri{integrare con nuovi lavori, sia AEG sia context-oriented. considerare se separare anche qui i due domini}

Automatic exploit generation (AEG) research challenge consists of automatically generating working exploits~\cite{avgerinos2014automatic}. 
This task requires technical skills and expertise in low-level languages to gain full control of the memory layout and CPU registers and attack low-level mechanisms (e.g., heap metadata and stack return addresses) not otherwise accessible through high-level programming languages. 
Given their recent advances, AI-code generators have become a new and attractive solution to help developers and security testers in this challenging task.
Although these solutions have shown high accuracy in the generation of software exploits, their robustness against new inputs has never been studied before.
Liguori \textit{et al.}~\cite{liguori2022can} released a dataset containing NL descriptions and assembly code extracted from software exploits. They performed an empirical analysis showing that NMT models can correctly generate assembly code snippets from NL and that in many cases can generate entire exploits with no errors. 
The authors extended the analysis to the generation of Python offensive code used to obfuscate software exploits from systems' protection mechanisms~\cite{liguori2021evil}.
Yang \textit{et al.}~\cite{yang2022dualsc} proposed a data-driven approach to software exploit generation and summarization as a dual learning problem. The approach exploits the symmetric structure between the two tasks via dual learning and uses a shallow Transformer model to learn them simultaneously.
Yang \textit{et al.}~\cite{yang2023exploitgen} proposed a novel template-augmented exploit code generation approach. The approach uses a rule-based template parser to generate augmented NL descriptions and uses a semantic attention layer to extract and calculate each layer’s representational information. %The authors show that the proposed approach outperforms the state-of-the-art baselines from the previous studies of automatic code generation.
Ruan \textit{et al.}~\cite{ruan2023prompt} proposed \textit{PT4Exploits}, an approach for software exploit generation via prompt tuning. They designed a prompt template to build the contextual relationship between English comment and the corresponding code snippet, simulating the pre-training stage of the model to take advantage of the prior knowledge distribution. 
Xu \textit{et al.}~\cite{xu2023autopwn} introduced an artifact-assisted AEG solution that automatically summarizes the exploit patterns from artifacts of known exploits and uses them to guide the generation of new exploits. The authors implemented \textit{AutoPwn}, an AEG system that automates the generation of heap exploits for Capture-The-Flag \textit{pwn} competitions. 
Recent work also explored the role of GPT-based models, including ChatGPT and Auto-GPT, in the offensive security domain. Botacin~\cite{botacin2023gpthreats} found that, by using these models, attackers can both create and deobfuscate malware by splitting the implementation of malicious behaviors into smaller building blocks.
Pa \textit{et al.}~\cite{pa2023attacker} and \cite{gupta2023chatgpt} proved the feasibility of generating malware and attack tools through the use of reverse psychology and \textit{jailbreak prompts}, i.e., maliciously crafted prompts able to bypass the ethical and privacy safeguards for abuse prevention of AI code generators like ChatGPT.
Gupta \textit{et al.}~\cite{gupta2023chatgpt} also examined the use of AI code generators to improve security measures, including cyber defense automation, reporting, threat intelligence, secure code generation and detection, and attack and malware detection.

%All the aforementioned works rely on the same high quality, manually curated \textit{Shellcode\_IA32} dataset~\cite{liguori2021shellcode_ia32}, which contains highly detailed descriptions of code and the corresponding assembly instructions. 
These studies do not take into account that, since developers have different levels of expertise and writing styles, code descriptions may be missing some information or reference previous instructions. The issue of inferring the context of the current sentence from the surrounding ones has been widely addressed for translation tasks between different natural languages. Proposed solutions vary from data-driven methods, to structural modifications to the model's architecture, or hybrid solutions. As for the former approach, previous studies concatenate previous and subsequent sentences, separated by a special token, to provide additional information to the model when translating the current phrase~\cite{DBLP:conf/discomt/TiedemannS17,scherrer-etal-2019-analysing, agrawal2018contextual}. Regarding architectural solutions, these include the integration of multiple encoders to encode not only the source-sentence, but also the context, i.e., the previous or next sentences~\cite{DBLP:conf/acl/LiLWJXZLL20,voita2018context}, or to encode the global context of the document~\cite{DBLP:conf/ijcai/ZhengYHCB20, DBLP:conf/emnlp/WangTWL17}.

These techniques have recently been applied also in the software engineering domain, in particular to aid the automatic generation of code comments and commit messages starting from programs.
Yu \textit{et al.}~\cite{xu2020towards} integrated local and class-level contextual information for code comment generation. They employ a local encoder, which extracts features from the target function, a global encoder, which exchanges information between all functions within the target class, and a decoder to aggregate local and global information. 
Wang \textit{et al.}~\cite{wang2021context} proposed a method to translate \textit{diffs}, i.e., the differences between two versions of code, that include both code changes and non-code changes into commit messages. They adopted retrieval-based solutions to retrieve the most similar commit from the training set to guide the commit message generation.

Our work builds upon previous studies to take advantage of contextual information coming from previous sentences also in the code generation domain. We use the concatenation of multiple intents to guide NMT models in generating more accurate assembly instructions even when the NL code descriptions provided by developers are missing important details.

\begin{comment}
### LAVORI su exploits generation

DualSC: Automatic Generation and Summarization of Shellcode via Transformer and Dual Learning
EVIL: Exploiting Software via Natural Language
Can we generate shellcodes via natural language? An empirical study
The shellcode generation
ExploitGen: Template-augmented exploit code generation based on CodeBERT

### LAVORI sul contesto

Context-Aware Retrieval-based Deep Commit Message Generation (architetturale?)
Towards Context-Aware Code Comment Generation (architetturale)

Neural Machine Translation with Extended Context (concatenazione)
Contextual Handling in Neural Machine Translation: Look Behind, Ahead and on Both Sides (both)
Exploiting Cross-Sentence Context for Neural Machine Translation (architetturale)
Does Multi-Encoder Help? A Case Study on Context-Aware Neural Machine Translation (both)
Context-Aware Neural Machine Translation Learns Anaphora Resolution (both)
Toward Making the Most of Context in Neural Machine Translation - BUONO PER RELATED WORK
One Type Context Is Not Enough: Global Context-aware Neural Machine Translation - 2022
\end{comment}

\section{Problem Statement}
\label{sec:problem}

AI code generation, powered by NMT models, represents a significant leap forward in bridging the gap between the conceptual intent expressed in NL and its translation into code. These advanced tools, trained on extensive datasets of intent-code pairs, have the capability to predict code snippets from a single NL description, encompassing everything from simple instructions to complex multi-line code segments.

%This breakthrough offers immense potential to streamline software development processes, enhance productivity, and make programming more accessible. 
However, the deployment of NMT models for code generation is not without its challenges. A primary concern is the inherent variability of NL descriptions, which can significantly impact the models' effectiveness.
Indeed, developers rarely provide exhaustive details in their code descriptions, often omitting information that they consider self-evident or redundant~\cite{improta2023enhancing}. This tendency towards brevity or assumed knowledge poses a significant challenge for NMT models, which rely on the completeness and clarity of the input to generate accurate code. The example provided in \tablename{}~\ref{tab:motivating_example} serves as a case in point, illustrating the discrepancies between human interpretation of NL instructions and the literal output of an NMT model.
In this example, the NMT model correctly interprets the first instruction to subtract a value from a byte in the ESI register. However, it falters when presented with a subsequent instruction that refers to the operation's outcome simply as ``the result.'' A human developer would instinctively understand this reference to imply the same register involved in the preceding operation, yet the model fails to make this connection, instead defaulting to a generic placeholder ``var'' due to the lack of explicit mention of the register's name.

\begin{table}[t]
\centering
\caption{Example of incorrect prediction. In \textcolor{red}{red}, the word implying the register name that the model fails to derive. }
\label{tab:motivating_example}
\small
\begin{tabular}
{>{\centering\arraybackslash}m{3cm} |   
>{\centering\arraybackslash}m{2cm} | >{\centering\arraybackslash}m{2cm}}
\toprule
\textbf{English Intent} & \textbf{Reference Code} & \textbf{Predicted Code}\\
\toprule
\textit{Subtract 8 from the current byte in ESI} & \texttt{sub byte [ESI], 8} & \texttt{sub byte [ESI], 8} \\
\midrule
\textit{Negate \textcolor{red}{the result}} & \texttt{not ESI} & \texttt{not \textcolor{red}{var}} \\
\bottomrule
\end{tabular}
\end{table}

This scenario underscores a critical limitation of current NMT models in code generation: their struggle with inferential reasoning and contextual understanding. For NMT models to be truly effective and reliable in a development setting, they must be capable of beyond-literal interpretation, grasping the unspoken implications of a given instruction based on its context within the broader scope of the code.

%There are several challenges in improving the ability of NMT models to generate a correct snippet in the absence of a detailed description of the desired code. The first objective is to assess whether and to what extent missing information from the current intent affects the code generation task. Then, to assess whether the model is able to gather the needed information from the previous intents and how far to look behind, i.e., how many previous intents to consider to support the current translation. Finally, another challenge to address is whether the model is capable of understanding when the contextual information is not useful, i.e., when previous intents do not contain details needed to translate the current source NL description.

To address these challenges, our study evaluates the impact of missing information on the code generation capabilities of NMT models, exploring their capacity to utilize context from preceding intents and discern the relevance of provided information. This involves examining the extent to which models can compensate for sparse or ambiguous NL descriptions by leveraging additional context and determining the optimal amount of contextual information that aids in accurate code translation without introducing confusion or inaccuracies.

By focusing on these areas, our research aims to shed light on the complex dynamics of AI-driven code generation and identify strategies for enhancing the contextual awareness and inferential capabilities of NMT models. The ultimate goal is to advance the state of AI code generation, ensuring that NMT models can effectively navigate the complexities of software development tasks and contribute to the creation of more intuitive, efficient, and accessible programming environments.

\section{Research Study}
\label{sec:research_study}
%To teach the model to understand how to look for the information that is missing from the current intent in the previous sentences, we extend the contextual information it has access to during training. We employ the \textit{concatenation} technique, i.e., we concatenate the previous intent or the two previous intents to the current source intent and associate to it the current target snippet. 
%To let the model discern the source sentence from the contextual information, we introduce an additional special token, i.e., a \texttt{\textit{\_BREAK}} token, to indicate the boundary between the intent to be translated and the previous ones. 
%Specifically, we design two strategies to provide contextual information:

To enhance the model's ability to decipher the complexity of NL descriptions for code generation, our study adopts a solution that enriches the contextual awareness (i.e., the \textit{contextual learning}) of NMT models during their training phase. %By integrating the technique of contextual concatenation, we aim to equip the models with a deeper understanding of the sequence and relevance of operations described in NL, thereby improving their ability to generate accurate and contextually appropriate code snippets.
Contextual learning, within the domain of AI code generation, refers to the capacity of NMT models to interpret and utilize the broader sequence of operations implied by a series of coding instructions~\cite{hwang2021contrastive}. Unlike traditional learning approaches that treat each instruction in isolation, contextual learning emphasizes understanding each piece of information within the continuum of previous and forthcoming instructions. This approach enables the model to make informed inferences about ambiguous or incomplete descriptions by leveraging the context provided by surrounding intents.

To enforce contextual learning in NMT models, we adopt a prompt engineering strategy that concatenates preceding intents to the current source intent. This method involves the strategic integration of additional context directly with the current instruction set, demarcated by a specially designed token, the \textit{\_BREAK} token~\cite{DBLP:conf/discomt/TiedemannS17}. This token serves a pivotal role in delineating the segments of concatenated intents, allowing the model to distinguish between the current actionable intent and its contextual backdrop.
Specifically, we design two strategies to provide contextual information:

\begin{itemize}
    \item \textbf{2to1 context:} we concatenate the previous intent to the current source intent, separated by the \texttt{\textit{\_BREAK}} token, to form a single NL code description as the input; the corresponding target code snippet represents the output. 
    \item \textbf{3to1 context:} we concatenate the two previous intents to the current source intent, separated by the \texttt{\textit{\_BREAK}} token, to form a single NL code description as the input; the corresponding target code snippet represents the output.
\end{itemize}

\begin{table}[t]
\centering
\caption{Examples of 2to1 and 3to1 context-aware intents. \textbf{Bold} refers to the current intent to be translated.}
\label{tab:context_examples}
\footnotesize
\begin{tabular}
{>{\centering\arraybackslash}m{1cm} |   
>{\centering\arraybackslash}m{4cm} |
>{\centering\arraybackslash}m{2.2cm}}
\toprule
\textbf{Context} & \textbf{English Intent} & \textbf{Reference Code Snippet}\\
\toprule
\multirow{3}{*}{\parbox{1cm}{\centering \textbf{2to1}}} & Clear the eax register \texttt{\textunderscore BREAK} \textbf{Move 22 into the lower byte} & \texttt{mov al,22} \\
\cmidrule{2-3}
& Move esi in eax \texttt{\textunderscore BREAK} \textbf{Increment it} & \texttt{inc eax} \\
\midrule
\multirow{4}{*}{\parbox{1cm}{\centering \textbf{3to1}}} & Subtract 8 from the current byte in esi \texttt{\textunderscore BREAK} Negate the result \texttt{\textunderscore BREAK} \textbf{Move eax in it} & \texttt{move byte[esi],eax} \\
\cmidrule{2-3}
& Move eax to edx \texttt{\textunderscore BREAK} Right shift the register by byte 16 \texttt{\textunderscore BREAK} \textbf{Add the result to eax} & \texttt{add eax, edx} \\
\bottomrule
\end{tabular}
\end{table}

%\cri{commenta tabella}
% tabella con esempio
% dataset vari, come sono costruiti?
\tablename{}~\ref{tab:context_examples} showcases examples of how the concatenation technique is applied to provide NMT models with contextual information, aiming to enhance their understanding and accuracy in code generation tasks. This technique is part of a strategy to improve model performance, particularly in scenarios where the NL description may lack detailed specificity. The examples are divided into two categories, based on the amount of preceding context provided: 2to1 and 3to1 contexts.

In the 2to1 context scenario, example \# 1 shows the intent ``Clear the eax register'' is followed by \textit{\_BREAK} and then the current task ``Move 22 into the lower byte.'' This provides a clear sequence of operations where the model is informed that the register is initially cleared before a new value is moved into it. The corresponding reference code snippet for the combined intents is \texttt{mov al, 22}, which directly corresponds to the current task, leveraging the context from the preceding intent.
Example \# 2 shows another instance combining ``Move esi in eax'' with the current intent ``Increment it,'' separated by the \textit{\_BREAK} token. This contextually rich input instructs the model that the ESI register's value has been moved to EAX before the increment operation, leading to the code \texttt{inc eax}. %The context helps clarify which register the ``Increment it'' command applies to.

The 3to1 context setup extends the concept by concatenating two previous intents to the current source intent, again using the \textit{\_BREAK} token for separation.
The first example starts with ``Subtract 8 from the current byte in esi,'' followed by ``Negate the result,'' and then the current intent ``Move eax in it.'' This sequence offers a comprehensive scenario where a subtraction is performed, the result is negated, and then a move operation is described, culminating in the generation of the code \texttt{move byte[esi],eax}. The detailed context aids the model in understanding the series of operations leading to the final action.
Example \# 2 presents a progression from ``Move eax to edx,'' through ``Right shift the register by byte 16,'' to the current intent ``Add the result to eax.'' This sequence provides a clear narrative of operations that the model can use to generate the corresponding code snippet \texttt{add eax, edx}, demonstrating the model's capacity to follow a multi-step process informed by the provided context.

%This solution does not lead to any modification to the model architecture itself, and it allows the NMT models to discover what kind of information is needed for the translation at a given time step.
%The self-attention mechanism of the Transformer architecture enables a simple strategy like concatenation of one or multiple contextual sentences to work better than more complex solutions such as multi-encoder-based approaches~\cite{agrawal2018contextual}.

Employing concatenation as a means to provide contextual information necessitates no alterations to the fundamental architecture of the NMT models. This simplicity in implementation, combined with the transformative potential in enhancing code generation accuracy, underscores the effectiveness of the solution. Moreover, the self-attention mechanism inherent in Transformer-based architectures is particularly well-suited to this strategy, enabling the models to dynamically adjust focus and relevance across the concatenated intents to derive a coherent and contextually informed output~\cite{agrawal2018contextual}.

%By leveraging the concatenation technique, this study aims to significantly advance the capability of NMT models in the realm of AI-driven code generation, particularly for applications requiring high levels of precision and contextual awareness, such as in the development of offensive security tools and protocols. Through this innovative approach, we endeavor to bridge the gap between the flexibility of natural language and the stringent demands of programming logic, paving the way for more intuitive and efficient software development methodologies.

\subsection{Research Questions}
We designed this research study with the aim of answering the following research questions (RQs):

%\textbf{R1:} Are NMT models robust to missing information in the NL descriptions?
\noindent
$\rhd$ \textbf{R1:} \textit{How do NMT models perform in generating offensive security code from NL descriptions when faced with missing information?}\\
This question seeks to understand the resilience of NMT models to incomplete or ambiguous NL descriptions, a common scenario in real-world applications. By evaluating the models' ability to fill in the gaps and make informed guesses about missing details, we aim to assess their practical utility in offensive security contexts, where the stakes of misinterpretation are high, and the cost of errors can be significant.

\noindent
$\rhd$ \textbf{R2:} \textit{Does contextual learning enhance the robustness of the NMT models in the generation of offensive security code?}\\
With RQ2, we delve into the potential of leveraging previous intents or contextual clues to improve model performance. This inquiry focuses on the models' ability to use additional, contextually relevant information to understand and accurately execute the current task. By exploring the impact of contextual learning, we seek to determine whether incorporating preceding intents as context can effectively mitigate the challenges posed by sparse or unclear NL descriptions.

\noindent
$\rhd$ \textbf{R3:} \textit{Does unnecessary information negatively impact the performance of the NMT models in the generation of offensive security code?}\\
The third question addresses the potential drawbacks of contextual information, particularly when it is irrelevant or superfluous. This aspect of the research is crucial for understanding the models' ability to discern and filter out unnecessary information, ensuring that their focus remains on relevant details crucial for the accurate generation of code. By investigating the impact of unnecessary context, we aim to reveal insights into how NMT models manage information overload and identify strategies for optimizing their training to improve focus and relevance in code generation tasks.

%\cri{invece di averla come RQ, facciamo un paragrafo a parte nella sezione sperimentale}
%\textbf{R4:} is our solution comparable to SOTA AI code generators in terms of performance?

\subsection{Code Generation Task}

To perform a rigorous evaluation of how the use of contextual information coming from previous intents affects the translation ability of the models, we follow the best practices in the field.
Hence, we support the models with \textit{data processing} operations. Data processing is an essential step to support the NMT models in the automatic code generation and refers to all the operations performed on the data used to train, validate and test the models.
%To support automatic code generation, neural machine translation is usually accompanied by data processing steps \citep{park2020decoding,8713737,oudah2019impact}. 
These operations strongly depend on the specific source and target languages to translate. 
The data processing steps are usually performed both before translation (\textit{pre-processing}), to train the NMT model and prepare the input data, and after translation (\textit{post-processing}), to improve the quality and the readability of the code in output.

First, we use a corpus to train the NMT models. The \textit{training data} is pre-processed before being used to feed the model. 
The pre-processing starts with the \textit{stopwords filtering}, i.e., we remove a set of custom-compiled words (e.g., \textit{the}, \textit{each}, \textit{onto}) from the intents to include only relevant data for machine translation. 
Next, we use a \textit{tokenizer} to break the intents into chunks of text containing space-separated words (i.e., the \textit{tokens}). 
To improve the performance of the machine translation~\cite{li2018named,modrzejewski2020incorporating,liguori2022can}, we \textit{standardize} the intents (i.e., we reduce the randomness of the NL descriptions) by using a \textit{named entity tagger}, which returns a dictionary of \textit{standardizable} tokens, such as specific values, label names, and parameters, extracted through regular expressions. We replace the selected tokens in every intent with ``\textit{var}\#", where \# denotes a number from $0$ to \textit{$|l|$}, and $|l|$ is the number of tokens to standardize.
Finally, the tokens are represented as real-valued vectors using \textit{word embedding}.  
The pre-processed data is used to feed the NMT model. 
Once the model is trained, we perform the code generation from NL. Therefore, when the model takes as inputs new intents from the \textit{test data} (i.e., data of the corpora not used in the training phase), it generates the corresponding code snippets based on the knowledge inferred during the training (\textit{model's prediction}).
As for the intents, also the code snippets predicted by the models are processed (\textit{post-processing}) to improve the quality and readability of the code. First, the dictionary of standardizable tokens is used in the \textit{de-standardization} process to replace all the ``\textit{var}\#" with the corresponding values, names, and parameters. 

Finally, the code snippets generated during the model's prediction are evaluated to assess the quality of the code generation task. The evaluation can be performed through output similarity metrics or manual analysis (human evaluation).
The former estimates the quality of the prediction by comparing the model's predictions with the ground truth reference in the test data, the latter, instead, assesses if the output predicted by the model is the correct translation of the NL intent into the generated code snippet.

\subsection{Fine-tuning Dataset}
\label{subsec:dataset}

%To fine-tune NMT models to generate offensive security code from NL descriptions, we carefully curated a dataset by drawing samples from two different sources. First, we collected $20$ real-world \textit{shellcodes} from reputable online databases and developer resources~\cite{exploitdb,shellstorm} and adopted in previous work to test models~\cite{liguori2021evil}.

To assess the impact of contextual learning in the generation of offensive security code from NL, we curated a dataset of real-world shellcodes from reputable online databases and developer resources~\cite{exploitdb,shellstorm}.
Since models require not just an understanding of individual code instructions but, crucially, how these instructions interact and depend on each other within a given context, our dataset consists of sequences of code that are contextually interconnected. 

\begin{table}[t]
\caption{The 20 shellcodes used to build the dataset for the experiments.}
\label{tab:test_set}
\centering
{\scriptsize
\begin{tabular}
{>{\centering\arraybackslash}m{0.5cm}| 
>{\tiny\arraybackslash}m{5.5cm} |
>{\centering\arraybackslash}m{1.5cm}
}
\toprule
\textbf{id} & \centering\textbf{URL} & \textbf{Lines of Code}\\
\toprule
1 & \url{https://www.exploit-db.com/shellcodes/47564} & 17 \\
2 & \url{https://www.exploit-db.com/shellcodes/47461} & 32 \\
3 & \url{https://www.exploit-db.com/shellcodes/46994} & 27 \\
4 & \url{https://www.exploit-db.com/shellcodes/46519} & 22 \\
5 & \url{https://www.exploit-db.com/shellcodes/46499} & 16 \\
6 & \url{https://www.exploit-db.com/shellcodes/46493} & 16 \\
7 & \url{https://www.exploit-db.com/shellcodes/45529} & 32 \\
8 & \url{https://www.exploit-db.com/shellcodes/43890} & 23 \\
9 & \url{https://www.exploit-db.com/shellcodes/37762} & 24 \\
10 & \url{https://www.exploit-db.com/shellcodes/37495} & 19 \\
11 & \url{https://www.exploit-db.com/shellcodes/43758} & 29 \\
12 & \url{https://www.exploit-db.com/shellcodes/43751} & 46 \\
13 & \url{https://rastating.github.io/creating-a-custom-shellcode-encoder/} & 27 \\
14 & \url{https://voidsec.com/slae-assignment-4-custom-shellcode-encoder/} & 18 \\
15 & \url{https://snowscan.io/custom-encoder/#} & 42 \\
16 & \url{https://github.com/Potato-Industries/custom-shellcode-encoder-decoder} & 19 \\
17 & \url{https://medium.com/@d338s1/shellcode-xor-encoder-decoder-d8360e41536f} & 33 \\
18 & \url{https://www.abatchy.com/2017/05/rot-n-shellcode-encoder-linux-x86} & 17 \\
19 & \url{https://xoban.info/blog/2018/12/08/shellcode-encoder-decoder/} & 24 \\
20 & \url{http://shell-storm.org/shellcode/files/shellcode-902.php} & 45 \\
 \bottomrule
\end{tabular}
}
\end{table}

In the domain of offensive code generation, shellcode generation represents a critical and widely studied topic~\cite{yang2023exploitgen,liguori2021evil,ruan2023prompt}. A shellcode is a list of machine code instructions to be loaded in a vulnerable application at runtime. The traditional way to develop shellcodes is to write them using the assembly language, and by using an assembler to turn them into \emph{opcodes} (operation codes, i.e., a machine language instruction in binary format, to be decoded and executed by the CPU)~\cite{foster2005sockets,megahed2018penetration}. Common objectives of shellcodes include spawning a system shell, killing or restarting other processes, causing a denial-of-service (e.g., a fork bomb), leaking secret data, etc.

To build the dataset, we first used $20$ real shellcodes adopted in previous work to test models~\cite{liguori2021evil}. The detailed source and number of lines of code (i.e., complexity), totaling $528$ lines of code, of the shellcodes used to build our dataset are shown in \tablename{}~\ref{tab:test_set}. These programs encompass a wide range of functionalities and complexities, ensuring a comprehensive evaluation of the models' code-generation capabilities.

To further extend our collection, we selected the $510$ \textit{multi-line snippets} from \textit{Shellcode\_IA32}~\cite{liguori2021shellcode_ia32}, a dataset comprising instructions in assembly language for IA-32 architectures from publicly-available security exploits and described in English.
A multi-line sample represents a set of instructions that it would be meaningless to consider as separate because of the strong contextual relation between them. Hereby, these samples are perfectly suited for our training objectives since they embody the contextual relationships between consecutive code snippets.
These lines correspond to code descriptions that generate multiple lines of shellcodes (separated by the newline character \textbackslash n). \tablename{}~\ref{tab:multiline_ex} shows an example drawn from the original dataset.
In order to ensure a diverse and non-redundant dataset, we carefully checked that there were no duplicates between the $20$ collected shellcodes and Shellcode\_IA32.
%We did not include the entire dataset because its NL descriptions are overly detailed for the purpose of the task at hand, which could bias the NMT models towards unrealistically precise inputs. To better simulate realistic and variable input scenarios, we intentionally modified the selected NL descriptions to be less accurate. This approach mimics the real-world inaccuracies often encountered in NL descriptions of code by different developers.

We opted not to include the entire Shellcode\_IA32 dataset, as it primarily comprises \textit{single-line} snippets that lack the contextual relationships (e.g., \texttt{mov eax, 1}). These snippets do not provide the sequential or logical linkages found in real-world programming tasks, which are crucial for testing and enhancing the contextual understanding capabilities of NMT models. Instead, we focused on using multi-line snippets that better represent the interconnected nature of offensive security code to build our dataset.

\begin{table}[t]
    \caption{Multi-line snippet from Shellcode\_IA32.}
    \label{tab:multiline_ex}
    \centering
    \small
    \begin{tabular}{m{8cm}}
        \toprule
        \textbf{Intent}: \textit{jump to the label recv http request if the contents of the \texttt{eax} register is not zero else subtract the value \texttt{0x6} from the contents of the \texttt{ecx} register}
        \\ \textbf{Multi-line Snippet}: \texttt{test eax, eax} \textbackslash n \texttt{jnz} \texttt{recv\_http request} \textbackslash n \texttt{sub ecx, 0x6} \\
        \bottomrule
    \end{tabular}
\end{table}

For the dataset's NL descriptions, we described the code collected from the shellcodes and selected samples from Shellcode\_IA32 to test the NMT models' capabilities across a wide array of offensive security coding challenges, yielding a total of $2,167$ pairs of NL descriptions and corresponding code snippets.
This process involved selecting instructions that either directly needed contextual understanding for code generation or, conversely, were deliberately disconnected to assess the models' ability to disregard irrelevant information. 
Specifically, we described the code snippets by using the following notations:

\noindent
$\blacksquare$ \textbf{No Context}: Instructions without added contextual information, forming the baseline for model performance assessment. This category includes 963 lines ($\sim 44\%$ of the dataset).\\
\textit{Example}: An intent such as "Increment the value in the register" with the code snippet \texttt{inc eax}, without any preceding context. This intent is presented without auxiliary information, challenging the model to deduce the target register based on common conventions or inferred knowledge.
    
\noindent
$\blacksquare$ \textbf{2to1 Context}: Incorporating the immediate preceding instruction to provide context, accounting for 360 lines ($\sim 17\%$ of the dataset).\\
\textit{Example}: "Clear the eax register \textbf{\textunderscore BREAK} Move 22 into the lower byte" translates to \texttt{mov al, 22}, where the context of clearing the register is crucial for understanding the operation.
    
\noindent
$\blacksquare$ \textbf{3to1 Context}: Extending context further by including the two instructions preceding the current one, comprising 238 lines ($\sim 11\%$ of the dataset).\\
\textit{Example}: "Subtract 8 from the current byte in esi \textbf{\textunderscore BREAK} Negate the result \textbf{\textunderscore BREAK} Move eax in it" results in \texttt{move byte[esi], eax}, showcasing the model's need to synthesize a broader sequence of operations.
    
\noindent
$\blacksquare$ \textbf{2to1 Unnecessary Context}: Featuring previous instructions that do not logically link to the current task, this scenario encompasses 303 lines ($sim 14\%$ of the dataset).\\
\textit{Example}: "Define the decode label \textbf{\textunderscore BREAK} Subtract 8 from the current byte of the shellcode" leads to \texttt{sub byte[esi], 8}, where the definition of a label is unrelated to the operation on the esi register.
    
\noindent
$\blacksquare$ \textbf{3to1 Unnecessary Context}: Similar to Unnecessary Context 2to1 but with two preceding instructions, also making up 303 lines ($\sim 14\%$ of the dataset).\\
\textit{Example}: "Increment edi \textbf{\textunderscore BREAK} Add 3 to al \textbf{\textunderscore BREAK} Jump short to switch" corresponds to \texttt{jmp short switch}, demonstrating the model's challenge in identifying relevant context from unrelated instructions.

The NL descriptions are tied to the relational structure of the code snippets derived from the shellcodes.
For the \textbf{No Context} scenario, where the objective is to evaluate the models' ability to generate code based solely on standalone instructions, we provided descriptions that omitted contextual dependencies. 
To better simulate realistic and variable input scenarios, we intentionally modified the original NL descriptions drawn from the Shellcode\_IA32 dataset, particularly precise and detailed, to be less accurate. This setup mimics the real-world inaccuracies often encountered in code descriptions provided by different developers.
%This setup simulates conditions where instructions are described without any prior context.
%
For contexts where the learning of relational dependencies was intended (\textbf{2to1 Context} and \textbf{3to1 Context}), we selected sequences of instructions that relied on each other to perform a coherent task. This required a description that could only be completed with the understanding of the previous one or two instructions, thus necessitating a detailed contextual description. Conversely, for scenarios meant to assess the models' handling of irrelevant information (\textbf{2to1 Unnecessary Context} and \textbf{3to1 Unnecessary Context}), we deliberately chose instructions that had no logical connection to each other. This setup aimed to test whether the models could effectively ignore context that does not contribute meaningfully to understanding the current task, thereby simulating conditions where contextual cues might mislead rather than aid in code generation. %These methodical choices ensure that each type of context—necessary and unnecessary—tests specific capabilities of the NMT models, reflecting realistic challenges they would face in operational environments.

The dataset's NL descriptions were independently crafted by three authors, all with a computer science background and expertise in assembly language and cybersecurity. The group included individuals with varying degrees of professional experience and educational qualifications. In particular, one Ph.D. student with a master's degree and two researchers with a Ph.D. in information technologies. The diversity and expertise of the authors ensured the reliability and variety of the dataset's NL descriptions construction process.

\section{Experimental Setup}
\label{sec:setup}
\begin{table*}[t]
\centering
\caption{Summary of data used in the training (Train), validation (Dev), and testing (Test) sets across different analyses.}
\label{tab:experimental_setup}
\scriptsize
\begin{tabular}{
>{\centering\arraybackslash}m{1.75cm} |
>{\centering\arraybackslash}m{4.75cm} |
>{\centering\arraybackslash}m{0.5cm} |
>{\centering\arraybackslash}m{1.1cm}
>{\centering\arraybackslash}m{1.1cm}
>{\centering\arraybackslash}m{1.1cm}
>{\centering\arraybackslash}m{1.1cm}
>{\centering\arraybackslash}m{1.1cm} |
>{\centering\arraybackslash}m{1.2cm}}
\toprule
\textbf{Analysis} & \textbf{Objective} & \textbf{Set} & \textbf{w/o Context} & \textbf{2to1 Context} & \textbf{3to1 Context} & \textbf{Unn. 2to1 Context} & \textbf{Unn. 3to1 Context} & \textbf{Total}\\
\toprule
\multirow{3}{1.5cm}{\textit{Missing Information (w/o context)}} & \multirow{3}{5cm}{Establishing a baseline for how well the models perform without any context to assess their innate ability to handle standalone instructions.} & Train & 770 (80\%) & - & - & - & - & 770 (80\%)\\
& & Dev & 96 (10\%) & - & - & - & - & 96 (10\%)\\
& & Test & 96 (10\%) & - & - & - & - & 96 (10\%)\\
\midrule
\multirow{3}{1.5cm}{\textit{Contextual Learning (2to1 context)}} & \multirow{3}{5cm}{Simulating a realistic coding scenario where previous information impacts the current operation.} & Train & 867 (90\%) & 180 (50\%) & - & - & - & 1047 (80\%)\\
& & Dev & 48 (5\%) & 90 (25\%) & - & - & - & 138 (10\%)\\
& & Test & 48 (5\%) & 90 (25\%) & - & - & - & 138 (10\%)\\
\midrule
\multirow{3}{1.5cm}{\textit{Contextual Learning (3to1 context)}} & \multirow{3}{4.75cm}{Emulating more complex coding scenarios to assess the model's ability in leveraging extended sequences of operations.} & Train & 867 (90\%) & - & 81 (34\%) & - & - & 948 (80\%)\\
& & Dev & 48 (5\%) & - & 79 (33\%) & - & - & 127 (10\%)\\
& & Test & 48 (5\%) & - & 79 (33\%) & - & - & 127 (10\%)\\
\midrule
\multirow{3}{2cm}{\textit{Unnecessary Information (2to1 Unn. Context)}} & \multirow{3}{5cm}{Determining the model's ability to discern and filter out unnecessary immediate context.} & Train & 867 (90\%) & 324 (90\%) & - & 103 (34\%) & - & 1293 (80\%) \\
& & Dev & 48 (5\%) & 18 (5\%) & - & 100 (33\%) & - & 166 (10\%)\\
& & Test & 48 (5\%) & 18 (5\%) & - & 100 (33\%) & - & 166 (10\%)\\
\midrule
\multirow{3}{2cm}{\textit{Unnecessary Information (3to1 Unn. Context)}} & \multirow{3}{5cm}{Evaluating the model's capacity to ignore multiple irrelevant preceding instructions.} & Train & 867 (90\%) & - & 214 (95\%) & - & 103 (34\%) & 1084 (80\%)\\
& & Dev & 48 (5\%) & - & 12 (5\%) & - & 100 (33\%) & 160 (10\%) \\
& & Test & 48 (5\%) & - & 12 (5\%) & - & 100 (33\%) & 160 (10\%)\\
\bottomrule
\end{tabular}
\end{table*}

In our experiments, we used a machine with a Debian-based distribution, 8 vCPU, 16 GB RAM, and one Nvidia T4 GPU. 

\tablename{}~\ref{tab:experimental_setup} summarizes the experimental setup across different analyses designed to evaluate the performance of NMT models in generating offensive security code from NL descriptions under various conditions. Each analysis aims to explore a specific aspect of the model's capabilities, such as handling missing information, leveraging contextual learning, and discerning unnecessary context. The dataset for each experiment is systematically divided into \textit{training} (i.e., the data used to fine-tune the models), \textit{validation} (i.e., the data used to tune the model's parameters), and \textit{test} sets (i.e., the data used to evaluate the model in the generation of the code starting from new NL descriptions), adhering to a common ratio of $80\%/10\%/10\%$~\cite{kim2018artificial,DBLP:conf/msr/MashhadiH21,liguori2021shellcode_ia32}, ensuring consistency and reliability in model evaluation.

We employed a consistent split ratio across all experiments to ensure that each model is fine-tuned, parameter-tuned, and evaluated under uniform conditions. The \textit{No Context} set serves as the foundation for comparing the impact of additional or unnecessary context, while the proportional inclusion of contextual scenarios across the train, validation, and test sets allows for a comprehensive assessment of how well NMT models learn from varying degrees of contextual information.
The availability of samples for each category also influenced our split. No context samples are more frequent than 2to1 context samples. Similarly, 2to1 context samples are more common than 3to1 context samples. This natural distribution affected how we allocated samples across the splits. Finally, we aimed to have a balanced sampling among training, validation, and test sets for contextual (necessary and unnecessary) information. We also aimed to have a test set size of close to 100 samples for each category to ensure a robust assessment.

Finally, we note that the size of the data used for the experiments is in line with other state-of-the-art corpora used to fine-tune models, which are relatively limited, i.e., in the order of one thousand samples~\cite{zhou2023lima}.
In fact, in state-of-the-art code generation, a model is not trained from scratch, but existing Large Language Models (that were already trained with millions of publicly available lines of code) are fine-tuned in a supervised way, to achieve transfer learning for the specific case (in our case, generating offensive code). 
We shared the dataset on GitHub~\footnote{\url{https://github.com/dessertlab/Software-Exploits-with-Contextual-Information}}.

%In the \textit{Missing Information} analysis, the focus is on establishing a baseline performance by assessing the model's ability to generate code without any contextual information. The entire set consists solely of instructions without context, split accordingly across the train, dev, and test sets.
%The \textit{Contextual Learning} analysis, both for 2to1 and 3to1 contexts, simulates realistic coding scenarios where preceding information (either one or two previous instructions) influences the generation of the current operation. This setup tests the models' ability to integrate and leverage additional context to improve code generation accuracy. 
%In analyses focused on \textit{Unnecessary Information} (both 2to1 and 3to1 Unnecessary Contexts), the objective is to evaluate the models' proficiency in filtering out irrelevant context. This scenario is particularly challenging as it introduces noise into the input data, testing the models' ability to maintain focus on the relevant parts of the input for accurate code generation. The unnecessary context data is introduced alongside the respective contextual data to observe any potential degradation in performance due to the inclusion of unrelated preceding instructions.

\subsection{NMT Models}
To perform the code generation task, we consider a comprehensive set of state-of-the-art models, which are described in the following.
The first three (i.e., CodeBERT, CodeT5+, and PLBart) are based on an \textit{encoder-decoder} architecture, where the input sequence is encoded into a context vector and then decoded to generate the output sequence. The latter (i.e., CodeGen, CodeGPT, and CodeParrot) are \textit{decoder-only} models, which read the input sequence and predict subsequent words one at a time, well-suited for generation tasks.

\noindent
%$\blacksquare$ \textbf{Seq2Seq} is a model that maps an input of sequence to an output of sequence. Similar to the encoder-decoder architecture with attention mechanism \cite{bahdanau2014neural}, we use a bi-directional LSTM as the encoder to transform an embedded intent sequence into a vector of hidden states with equal length. We implement the Seq2Seq model using \textit{xnmt}~\cite{neubig-etal-2018-xnmt}.  We use an Adam optimizer \cite{kingma2015adam} with $\beta_1=0.9$ and $\beta_2=0.999$, while the learning rate $\alpha$ is set to $0.001$. We set all the remaining hyper-parameters in a basic configuration: layer dimension = $512$, layers = $1$, epochs = $200$, beam size = $5$.

\noindent
$\blacksquare$ \textbf{CodeBERT}~\cite{feng2020codebert} is a large multi-layer bidirectional Transformer architecture~\cite{vaswani2017attention} pre-trained on millions of lines of code across six different programming languages. 
%Like Seq2Seq, the Transformer architecture is made up of encoders and decoders. %CodeBERT has 12 stacked encoders and 6 stacked decoders. 
Our implementation uses an encoder-decoder framework where the encoder is initialized to the pre-trained CodeBERT weights, and the decoder is a transformer decoder, composed of $ 6$ stacked layers. The encoder follows the RoBERTa architecture~\cite{DBLP:journals/corr/abs-1907-11692}, with $12$ attention heads,  hidden layer dimension of $768$, $12$ encoder layers, and $514$ for the size of position embeddings. We set the learning rate $\alpha = 0.00005$, batch size = $32$, and beam size = $10$.

\noindent
$\blacksquare$ \textbf{CodeT5+}~\cite{wang2023codet5+} is a new family of Transformer models pre-trained with a diverse set of pretraining tasks including causal language modeling, contrastive learning, and text-code matching to learn rich representations from both unimodal code data and bimodal code-text data. 
We utilize the variant with model size $220M$, which is trained from scratch following T5’s architecture~\cite{DBLP:journals/jmlr/RaffelSRLNMZLL20}. It has an encoder-decoder architecture with $12$ decoder layers, each with $12$ attention heads and hidden layer dimension of $768$, and $512$ for the size of position embeddings. We set the learning rate $\alpha = 0.00005$, batch size = $16$, and beam size = $10$.

\noindent
$\blacksquare$ \textbf{PLBart}~\cite{ahmad2021unified} is a multilingual encoder-decoder (sequence-to-sequence) model primarily intended for code-to-text, text-to-code, code-to-code tasks. The model is pre-trained on a large collection of Java and Python functions and natural language descriptions collected from GitHub and StackOverflow.
We use the PLBart-large architecture with $12$ encoder layers and $12$ decoder layers, each with $16$ attention heads. We set the learning rate $\alpha = 0.00005$, batch size=$16$, and beam size=$10$.

\noindent
$\blacksquare$ \textbf{CodeGen}~\cite{codegen}, is an autoregressive language model for program synthesis with an architecture that follows a standard transformer decoder with left-to-right causal masking. %The family of CodeGen models is trained in various sizes, including 350M, 2.7B, 6.1B, and 16.1B, and utilizes various datasets. 
Specifically, we leverage \textit{CodeGen-350M-Multi}, initialized from CodeGen-NL and further pre-trained on BigQuery \cite{codegen},  a large-scale dataset of multiple programming languages from GitHub repositories, which consists of 119.2B tokens and includes C, C++, Go, Java, JavaScript, and Python.

\noindent
$\blacksquare$ \textbf{CodeGPT}~\cite{lu2021codexglue}, is a Transformer-based language model pre-trained on millions of Python functions and Java methods. The decoder-only architecture consists of 12 layers of Transformer decoders with $124M$ parameters. We adopted the \textit{CodeGPT-small-py-adaptedGPT2} version, which has the same GPT-2 vocabulary and natural language understanding ability to support text-to-code generation tasks.
We followed previous work for the implementation~\cite{lu2021codexglue}.

\noindent 
$\blacksquare$ \textbf{CodeParrot}~\cite{CodeParrot} is a GPT-2 model with BPE tokenizer trained on Python code from the training split of the data, and a context length of 1024. This model was released as an educational tool for training large language models from scratch on code, with detailed tutorials and descriptions of the training process. We adopted the \textit{codeparrot-small} version with $110M$ parameters.

During data pre-processing, we tokenize the NL intents using the \textit{nltk word tokenizer}~\cite{bird2006nltk} and code snippets using the Python \textit{tokenize} package~\cite{tokenize}. 
We use \emph{spaCy}, an open-source, NL processing library written in Python and Cython~\cite{spacy}, to implement the named entity tagger for the standardization of the NL intents.

\subsection{Metrics}
Following best practices in the field,  we adopted \emph{output similarity metrics} to assess the performance of the models. These metrics, which compare the generated code with the code from the ground truth, are widely used to assess the performance of AI generators in many code generation tasks~\cite{LIGUORI2023120073}, including the generation of assembly code for security contexts~\cite{yang2022dualsc,yang2023exploitgen,ruan2023prompt,liguori2021evil,liguori2022can}. In particular, we adopted the following metrics:

%\item \textbf{Compilation Accuracy (CA)}. It indicates whether each code snippet produced by the model is compilable according to the syntax rules of the target language. CA value is either $1$, when the snippet's syntax is correct, or $0$ otherwise. To compute the \textit{compilation accuracy}, we used the \textit{Netwide Assembler} (NASM) assembler~\citep{nasm}.
%\item \textbf{Bilingual Evaluation Understudy (BLEU) score}~\cite{papineni2002bleu}. It measures the degree of n-gram overlapping between the string of each code snippet produced by the model and the reference. This metric also takes into account a \textit{brevity penalty} to penalize predictions shorter than the references. BLEU value ranges between $0$ and $1$, with higher scores corresponding to a better quality of the prediction.  Similar to previous studies, we use the BLEU-4 score (i.e., we set $n=4$). We implemented BLEU score computation employing the \texttt{bleu\_score} module contained in the open-source Python suite Natural Language Toolkit (NLTK)~\cite{bleu}. 
%\item \textbf{SacreBLEU}~\cite{post-2018-call}. This is a different implementation of the BLEU score which differs from the traditional one because it uses different tokenization techniques. We used the implementation available on Hugging Face~\cite{SacreBLEU}
\noindent
$\blacksquare$ \textbf{Exact Match accuracy (EM)}. It indicates whether each code snippet produced by the model perfectly matches the reference. EM value is $1$ when there is an exact match, $0$ otherwise. To compute the exact match, we used a simple Python string comparison.

\noindent
$\blacksquare$ \textbf{Edit Distance (ED)}. It measures the \textit{edit distance} between two strings, i.e., the minimum number of operations on single characters required to make each code snippet produced by the model equal to the reference. ED value ranges between $0$ and $1$, with higher scores corresponding to smaller distances. For the edit distance, we adopted the Python library \texttt{pylcs}~\cite{pylcs}. 

\noindent
$\blacksquare$ \textbf{METEOR}\cite{10.5555/1626355.1626389}. It measures the \textit{alignment} between each code snippet produced by the model and the reference. The alignment is defined as a mapping between unigrams (i.e., $1$-gram), such that every unigram in each string maps to zero or one unigram in the other string, and no unigrams in the same string. METEOR value ranges between $0$ and $1$, with higher scores corresponding to greater alignment between strings. To calculate the METEOR metric, we relied on the Python library \texttt{evaluate} by HuggingFace~\cite{meteor}.

\noindent $\blacksquare$  \textbf{ROUGE-L}. It is a metric based on the longest common subsequence (LCS) between the model's output and the reference, i.e. the longest sequence of words (not necessarily consecutive, but still in order) that is shared between both. The metric ranges between $0$ (perfect mismatch) and $1$ (perfect matching). We computed the ROUGE-L metric using the Python package \texttt{rouge}~\cite{rouge}.

\section{Experimental Evaluation}
\label{sec:experimental}
This section describes the experimental setup and results obtained from evaluating several NMT models on the task of generating assembly shellcodes from NL descriptions. The experiments were designed to assess the models' ability to handle missing information, leverage extended context effectively, and discern and utilize unnecessary information.

%5) CREATO TEST SET CON 20 SHELLCODE, SENZA CONTESTO, TRAIN ESTESO	
\subsection{Missing Information}

First, we evaluated the performance of NMT models when generating offensive security code from NL descriptions with no additional contextual information (see \tablename{}~\ref{tab:experimental_setup}).  This scenario simulates real-world conditions where developers may provide incomplete or vague descriptions due to oversight or assumption of implicit knowledge. Understanding how NMT models cope with such missing information is crucial for assessing their practical applicability in automated code generation tasks.
\tablename{}~\ref{tab:missing_context} shows the results.

\begin{table}[t]
\centering
\caption{Model performance on the generation task with missing context.}
\label{tab:missing_context}
\footnotesize
\begin{tabular}{
>{\centering\arraybackslash}m{1.5cm} |
%>{\centering\arraybackslash}m{1cm}
>{\centering\arraybackslash}m{1cm}
>{\centering\arraybackslash}m{1cm}
>{\centering\arraybackslash}m{1.25cm}
>{\centering\arraybackslash}m{1.5cm}}
\toprule
\textbf{Model} %& \textbf{BLEU-4}
& \textbf{EM} & \textbf{ED} & \textbf{METEOR} & \textbf{ROUGE-L} \\ \toprule
\textit{CodeBERT}   %& 80.34
& 45.99 & 77.30 & 67.41 & 65.75 \\
\textit{CodeT5+}    %& 75.87
& 59.35 & 81.68 & 74.87 & 73.44 \\
\textit{PLBart}     %& 1.14
& 7.44  & 32.59 & 21.41 & 27.04 \\
\textit{CodeGen}    %& 23.73
& 30.92 & 60.97 & 52.72 & 48.76 \\
\textit{CodeGPT}    %& 19.00
& 16.60 & 46.95 & 38.60 & 34.17 \\
\textit{CodeParrot} %& 17.23
& 21.76 & 53.15 & 43.78 & 40.71 \\ \midrule
\textbf{All Models} %& 36.22
& 30.34 & 58.77 & 49.80 & 48.31 \\
\bottomrule
\end{tabular}
\end{table}

CodeBERT and CodeT5+ exhibit superior performance across all metrics, with CodeT5+ showing particularly high scores in EM (59.35), METEOR (74.87), and ROUGE-L (73.44). CodeBERT also performs well, with its best score in METEOR (67.41). This indicates a strong capability of the models in accurately translating NL descriptions to code, suggesting a robust understanding of the language and the task requirements even in the absence of context. 
PLBart struggles significantly in this setup, with much lower scores across the board, particularly in EM (7.44) and ROUGE-L (27.04). This may indicate difficulties in capturing the essence of the NL descriptions and translating them into accurate code snippets without contextual cues.
CodeGen, CodeGPT, and CodeParrot display moderate performances, with CodeGen performing relatively better among the three, especially in terms of EM (30.92) and ED (60.97). These models seem to have a moderate capability in understanding and generating code from NL descriptions, with varying degrees of success in handling the missing context.
The Average scores across models (EM: 30.34, ED: 58.77, METEOR: 49.80, ROUGE-L: 48.31) reflect a collective moderate proficiency in dealing with NL descriptions devoid of additional context. This underscores a variation in how different models process and generate code based solely on the information contained within the NL instructions.

To provide context for the evaluation, it's instrumental to consider findings from previous research that tackled a similar task but with more detailed NL descriptions~\cite{yang2023exploitgen}. Specifically, prior work reported EM scores of 48.52 for CodeGPT and 51.80 for CodeBERT, indicating a benchmark for models generating assembly code under conditions of richer linguistic input.
In our experiments, CodeGPT and CodeBERT achieved EM scores of 16.60 and 45.99, respectively, when operating without contextual information. This represents a notable decline for CodeGPT compared to the previous EM score of 48.52 and a slight underperformance for CodeBERT relative to its previous benchmark of 51.80.
The discrepancy in performance between the current and previous studies, particularly for CodeGPT, underscores the impact of missing information on model output. %CodeBERT's resilience, albeit with a minor drop in performance, suggests a certain degree of robustness in handling NL descriptions even when they lack detail.

\begin{mybox}{\parbox{8cm}{RQ1: How do NMT models perform in generating offensive security code from NL descriptions when faced with missing information?}}
NMT models exhibit varied levels of proficiency in generating offensive security code from less detailed NL descriptions. The comparative analysis reveals a marked impact of missing information on the accuracy of code generation, particularly for CodeGPT, which saw a significant performance dip compared to its prior achievements. Meanwhile, CodeBERT demonstrates a relative steadiness, albeit with room for improvement to match its earlier performance.
This juxtaposition highlights the critical role of detailed NL descriptions in achieving high accuracy in code generation and also points to the essential need for enhancing NMT models' capabilities in dealing with sparse information. It emphasizes an ongoing research imperative to develop models that can more effectively bridge the gap between minimalistic NL instructions and the precise requirements of low-level programming tasks, particularly in the contextually rich domain of offensive security.
\end{mybox}

%7) CREATO TEST SET CON 20 SHELLCODE, CONTESTO AD 1, TRAIN ESTESO					

%9) CREATO TEST SET CON 20 SHELLCODE, CONTESTO AD 2, TRAIN ESTESO		
\subsection{Contextual Learning}

\begin{table}[t]
\centering
\caption{Comparison of models' performance with 2to1 and 3to1 Context.}
\label{tab:extended_context}
\footnotesize
\begin{tabular}{
>{\centering\arraybackslash}m{1.25cm} |
>{\centering\arraybackslash}m{1.25cm} |
%>{\centering\arraybackslash}m{1cm}
>{\centering\arraybackslash}m{0.75cm}
>{\centering\arraybackslash}m{0.75cm}
>{\centering\arraybackslash}m{1cm}
>{\centering\arraybackslash}m{1cm}}
\toprule
\textbf{Model} & \textbf{Context Type} %& \textbf{BLEU-4}
& \textbf{EM} & \textbf{ED} & \textbf{METEOR} & \textbf{ROUGE-L} \\ \toprule
\multirow{2}{*}{\textit{CodeBERT}}   & 2to1 %& 67.10
& 61.07 & 80.77 & 74.30 & 72.29 \\
  & 3to1 %& 74.31 
  & 47.90 & 75.34 & 67.19 & 65.84 \\\midrule
\multirow{2}{*}{\textit{CodeT5}+}    & 2to1 %& 70.37
& 62.40 & 82.72 & 77.58 & 75.50 \\
   & 3to1 %& 61.74
   & 62.79 & 81.70 & 77.06 & 75.14 \\\midrule
\multirow{2}{*}{\textit{PLBart}}     & 2to1 %& 6.27
& 13.55 & 35.99 & 29.36 & 33.21 \\
     & 3to1 %& 4.91
     & 10.31 & 24.25 & 21.63 & 29.50 \\\midrule
\multirow{2}{*}{\textit{CodeGen}}    & 2to1 %& 20.61
& 43.70 & 66.15 & 60.87 & 55.91 \\
    & 3to1 %& 16.60
    & 36.26 & 62.07 & 53.91 & 51.43 \\\midrule
\multirow{2}{*}{\textit{CodeGPT}}    & 2to1 %& 20.28
& 25.38 & 53.71 & 44.59 & 41.14 \\
    & 3to1 %& 18.86
    & 22.14 & 50.83 & 41.20 & 37.96 \\\midrule
\multirow{2}{*}{\textit{CodeParrot}} & 2to1 %& 16.02
& 27.67 & 57.98 & 48.56 & 45.24 \\
 & 3to1 %& 12.75
 & 20.99 & 51.52 & 42.20 & 37.16 \\\midrule
\multirow{2}{*}{\textbf{All Models}} & 2to1 %& 33.44	
 & 38.96	& 62.89	& 55.88	& 53.88\\
 & 3to1 %& 31.53	
 & 33.40	& 57.62	& 50.53	& 49.51 \\
\bottomrule
\end{tabular}
\end{table}

Then, we focused on analysing the impact of contextual learning on the performance of NMT models in generating assembly shellcodes from NL descriptions. The extended context was provided in two forms: one additional previous intent (2to1 context) and two additional previous intents (3to1 context) by using the 2to1 Context Test and the 3to1 Context Test, respectively (see \tablename{}~\ref{tab:experimental_setup}).
This investigation is particularly important for determining how much contextual information can effectively aid in enhancing model performance, especially in comparison to baseline performance without any extended context.
\tablename{}~\ref{tab:extended_context} shows the results of the analysis.

The results show that all models improved performance in the 2to1 context compared to the no-context scenario, with CodeT5+ (EM: 62.40) and CodeBERT (EM: 61.07) leading the pack. This indicates a substantial benefit from including immediate preceding context, as it likely provides crucial information missing from the standalone instructions.
The average EM score increased by 8.62 points, METEOR by 6.08, and ROUGE-L by 5.57, underscoring the utility of adding a single contextual sentence in bridging the gap between the NL descriptions and the required code outputs.

In the 3to1 Context experiments, while CodeT5+ again excels (EM: 62.79), other models like CodeBERT and PLBart show decreased performance compared to the 2to1 context, suggesting a potential overload or diminishing returns from additional context for some models. Indeed, the transition from 2to1 to 3to1 context shows an overall decrease in performance averages across metrics, with a notable drop in EM (-5.56) and METEOR (-5.35). This implies that while some context is beneficial, too much can confuse models or dilute the relevant information.
Despite mixed results when transitioning from 2to1 to 3to1 contexts, comparing the 3to1 context results to the no-context scenario reveals an overall enhancement in model performance. The average EM score shows an improvement (3.06 points increase), indicating that even with potential issues of information overload, the inclusion of an extended context offers a net benefit over standalone instructions.

Previous work involving detailed NL descriptions and the use of models like CodeGPT and CodeBERT achieved EM scores of 48.52 and 51.80, respectively. When considering our findings in the 2to1 context experiment, where CodeBERT and CodeT5+ exhibit significant performance boosts (61.07 and 62.40 in EM, respectively), it suggests a parallel in how contextual information can supplement detailed NL descriptions in enhancing model output. The ability of models to perform comparably or even better with added context against a backdrop of detailed descriptions underscores the value of contextual learning in bridging the gap between NL instructions and code generation tasks.

\begin{mybox}{\parbox{8cm}{RQ2: Does contextual learning enhance the robustness of the NMT models in the generation of offensive security code?}} 
Contextual learning enhances the robustness of NMT models in generating offensive security codes, affirming the positive impact of incorporating contextual information. Specifically, the addition of a single preceding instruction (2to1 context) markedly improves model performance across various metrics, demonstrating that even minimal context can significantly aid models in understanding code generation tasks more accurately.
Comparing the 3to1 context to the no-context scenario reveals that extended context still offers benefits, with all models performing better than without any context. This further corroborates the value of contextual information in enhancing model performance. However, the slight decline in metrics when moving from a 2to1 to a 3to1 context suggests an optimal level of contextual information beyond which the effectiveness of additional context may diminish or even detract from model performance. %This highlights the need for a balanced approach to contextual learning, where the benefits of additional context are weighed against the potential for information overload.
\end{mybox}

\subsection{Unnecessary Information}

%10) TEST SET CON UNNECESSARY CONTEXT AD 1	\\	
%11) TEST SET CON CONTEXT 1, TRAIN E DEV CON UNNECESSARY CONTEXT 1					

%12) TEST SET CON UNNECESSARY CONTEXT AD 1 E 2 \\
%12) TEST SET CON CONTEXT 2, TRAIN E DEV CON UNNECESSARY CONTEXT AD 1 E 2	

\begin{table}[t]
\centering
\caption{Comparison of models' performance with 2to1 and 3to1 Unnecessary Context.}
\label{tab:unnecessary_context_1_2}
\footnotesize
\begin{tabular}{
>{\centering\arraybackslash}m{1.25cm} |
>{\centering\arraybackslash}m{1.25cm} |
%>{\centering\arraybackslash}m{1cm}
>{\centering\arraybackslash}m{0.75cm}
>{\centering\arraybackslash}m{0.75cm}
>{\centering\arraybackslash}m{1cm}
>{\centering\arraybackslash}m{1cm}}
\toprule
\textbf{Model} & \textbf{Context Type} %& \textbf{BLEU-4}
& \textbf{EM} & \textbf{ED} & \textbf{METEOR} & \textbf{ROUGE-L} \\ \toprule
\multirow{2}{*}{\textit{CodeBERT}}   & 2to1 %& 50.38
& 30.15 & 63.51 & 52.03 & 48.71 \\
           & 3to1 %& 71.68
           & 35.69 & 65.08 & 55.68 & 52.98 \\\midrule
\multirow{2}{*}{\textit{CodeT5+}}    & 2to1 %& 42.48
& 45.42 & 71.78 & 61.99 & 60.75 \\
           & 3to1 %& 80.69
           & 47.33 & 74.24 & 63.59 & 61.75 \\\midrule
\multirow{2}{*}{\textit{PLBart}}     & 2to1 %& 7.39
& 16.98 & 36.37 & 29.41 & 32.50 \\
           & 3to1 %& 0.85
           & 7.25  & 26.37 & 17.78 & 21.28 \\\midrule
\multirow{2}{*}{\textit{CodeGen}}    & 2to1 %& 22.62
& 61.83 & 81.74 & 67.92 & 71.11 \\
           & 3to1 %& 20.14
           & 50.76 & 73.37 & 59.78 & 59.51 \\\midrule
\multirow{2}{*}{\textit{CodeGPT}}    & 2to1 %& 23.90
& 46.76 & 72.79 & 57.35 & 59.47 \\
           & 3to1 %& 21.07
           & 37.02 & 66.37 & 49.70 & 49.87 \\\midrule
\multirow{2}{*}{\textit{CodeParrot}} & 2to1 %& 19.68
& 54.39 & 77.22 & 63.25 & 64.97 \\
           & 3to1 %& 17.23
           & 34.16 & 66.03 & 49.70 & 47.56 \\\midrule
\multirow{2}{*}{\textbf{All Models}} & 2to1 %& 27.74
& 42.59	& 67.24	& 55.33	& 56.25\\
 & 3to1 %& 35.28	
 & 35.37	& 61.91	& 49.37	& 48.83\\
\bottomrule
\end{tabular}
\end{table}

Finally, we explored the impact of incorporating irrelevant contextual information (Unnecessary Context 2to1 and 3to1) on the performance of NMT models in generating offensive security code (see \tablename{}~\ref{tab:experimental_setup}). This setup allows us to investigate the ability of the models to generate precise code in the presence of unnecessary context.
\tablename{}~\ref{tab:unnecessary_context_1_2} shows the results.

The overall performance increased by 12.25, 8.46, 5.53, and 7.94 points for EM, ED, METEOR, and ROUGE-L, respectively, over the no-context baseline, indicating that models are capable of filtering out irrelevant information to some extent and still improving upon the no-context scenario.

Decoder-only models like CodeGen, CodeGPT, and CodeParrot showed significant improvements in EM scores (61.83, 46.76, and 54.39, respectively), even outperforming their results in the 2to1 context experiment. This suggests that, based on model architecture, the presence of unnecessary context does not hinder—and may even aid—their code generation capabilities. 
These models rely on causal attention, where each generated token attends to all previous tokens. This allows the model to weigh each input token’s importance based on its contribution to the output sequence. When the previous sentence is unrelated, such is in this analysis, it likely receives lower attention weights during the generation process, allowing the model to focus more on the relevant target description. %Conversely, when the previous sentence is related, the attention mechanism might distribute focus across both the previous and target sentences. This can lead to the blending of information, where the model may consider context from the related sentence that is not directly useful for the specific code to be generated.

The transition to unnecessary 3to1 context shows a notable impact on model performance. While CodeT5+ and CodeGen maintain relatively high performance, there's a notable decrease across all models when directly compared to the unnecessary 2to1 context setup. This decrease underscores the models' challenges in managing more complex distractions.
Despite the presence of unnecessary information, models generally maintain or slightly improve performance compared to the no-context baseline, albeit with some performance reduction compared to the 2to1 necessary context setup.

\begin{mybox}{\parbox{8cm}{RQ3: Does unnecessary information negatively impact the performance of the NMT models in the generation of offensive security code?}} 
Unnecessary information does not universally negatively impact the performance of NMT models in generating offensive security code. In the unnecessary 2to1 context setup, models demonstrate a remarkable ability to disregard irrelevant information, with decoder-only models even showing improved performance over both the no-context and necessary 2to1 context scenarios. This improvement suggests that models can extract useful patterns or ignore distractions, focusing on the task-relevant aspects of the input, also based on their architecture.
However, as the complexity of unnecessary context increases (unnecessary 3to1), we observe a general performance decrease compared to unnecessary 2to1 context, highlighting a limit to the models' ability to filter out noise. This performance decline, though slight, emphasizes the challenge of managing more extensive unrelated information, which can obscure relevant details and impede accurate code generation.
%Thus, while unnecessary information does not invariably degrade model performance and may even offer unexpected benefits, its impact becomes more pronounced with increasing complexity. These findings suggest a robustness in current NMT models to some degree of irrelevant context, but also point to the importance of optimizing the quantity and quality of contextual information provided during model training and inference to maximize accuracy and efficiency in code generation tasks.
\end{mybox}

%\section{Discussion}
%\label{sec:discussion}
%\input{tex/discussion}

\section{Threats to Validity}
\label{sec:threats}
\noindent
\textbf{Models Selection.} The external validity of the study might be impacted by the choice of models. To mitigate this, we carefully selected models with distinct architectures and capabilities, ensuring a representation of current advancements in the field~\cite{lu2021codexglue, wei2023copiloting, tipirneni2022structcoder,cotroneo2024vulnerabilities}. 
This careful selection aims to ensure that our findings reflect broader trends in NMT model performance for code generation tasks.
Moreover, we did not consider public AI models such as GitHub Copilot and OpenAI ChatGPT because they impose restrictions on malicious uses and, as a consequence, they often ``refuse'' to generate offensive code, even if it is used for research or testing activities. Also, since both attackers and defenders need to avoid leaking their techniques and tactics to their counterparts (OPSEC), we consider the case of an attacker or defender that builds her own AI code generator by fine-tuning a model on a dataset of offensive code, thus circumventing usage policies of public AI code generators. %Therefore, we fine-tuned six standard architectures, representative of the SOTA, on a corpus of offensive code.

\noindent
\textbf{Evaluation Metrics.} 
The reliance on output similarity metrics, although representing the most common solution in the field, poses a potential threat to construct validity, as these metrics may not fully encapsulate the correctness of the generated assembly code~\cite{cotroneo2024automating}. To address this issue, our evaluation strategy encompassed a comprehensive suite of metrics, each offering unique insights into the models' performance. By considering multiple metrics and aligning with common practices in code generation evaluation, we provided a well-rounded assessment. No single metric is perfect, but analyzing them collectively allows for a comprehensive evaluation of the code.

\noindent
\textbf{Dataset Construction and NL Description.} %Our dataset, while derived from reputable sources, may not cover all possible scenarios, coding styles, or exploit techniques found in offensive security. 
The selection of $20$ real shellcodes and the multi-line samples from Shellcode\_IA32 to build the dataset aimed to provide a broad overview of common tasks and challenges in this domain. However, the inherent variability in exploit development and the continuous evolution of offensive techniques mean that no dataset can fully encapsulate the entire scope of offensive coding. To mitigate this threat, we ensured that the chosen shellcodes span a range of functionalities, complexities, and objectives, aiming for a balanced representation of real-world coding tasks in offensive security. 
%\textcolor{blue}{Moreover, we did not include the entire Shellcode\_IA32 dataset for our experimental evaluation, as it primarily comprises \textit{single-line} snippets that lack the contextual relationships crucial for testing and enhancing the contextual understanding capabilities of NMT models. Instead, we focused on using multi-line snippets that better represent the interconnected nature of offensive security code to build our dataset.}
Regarding the NL descriptions of the shellcodes, we aimed to mimic the variability in detail and specificity that might be encountered in real-world scenarios. Nevertheless, we acknowledge that the manual description of shellcodes introduces subjectivity and could potentially influence the models' performance. To minimize bias and ensure consistency, multiple authors independently described different samples of the dataset, and, where available, we used developers' original comments as NL descriptions. This multi-faceted approach to dataset annotation seeks to capture a realistic range of expression and technical detail, enhancing the external validity of our findings.

%\noindent
%\textbf{Dataset:} This work targets the assessment of AI-based solutions to generate offensive code for software exploits. The dataset used for our experiments fits perfectly with the scope of this work since it is the largest collection of offensive code available to date for code generation. This manually curated dataset contains high-quality and detailed descriptions of code, that are often not available in larger corpora for code generation. Indeed, the dataset provides NL descriptions both at the block and statement levels that are closer to the descriptions needed by the NMT models for complex programming tasks.  Moreover, we notice that the size of our dataset is in line with other state-of-the-art corpora used to fine-tune ML models. In fact, in state-of-the-art code generation, a model is not trained from scratch, but existing Large Language Models (that were already trained with millions of publicly available lines of code) are fine-tuned in a supervised way, to achieve transfer learning for the specific case (in our case, generating offensive code). Typically, the datasets for fine-tuning are relatively limited, in the order of one thousand samples~\cite{zhou2023lima}.

\section{Conclusion}
\label{sec:conclusion}
This work presented a comprehensive investigation into the capabilities of NMT models in the domain of offensive security code generation. Through a series of designed experiments, we investigated the complexities surrounding the models' interaction with missing information, their utilization of contextual learning, and their resilience in the face of unnecessary context. 

Our results showed that contextual information substantially improved model performance, highlighting the value of contextual information in dealing with undetailed NL descriptions. However, the results also showed the diminishing returns associated with extensive context, pointing to the existence of an optimal context threshold that maximizes model performance.
In scenarios involving unnecessary context, our experiments demonstrated the models' ability to filter out irrelevant information, maintaining high performance in the code generation task.

\section*{Acknowledgment}
This work has been partially supported by the MUR PRIN 2022 program, project \textit{FLEGREA}, CUP E53D23007950001, the \textit{IDA—Information Disorder Awareness} Project funded by the European Union-Next Generation EU within the SERICS Program through the MUR National Recovery and Resilience Plan under Grant PE00000014, the \textit{SERENA-IIoT} project funded by MUR (Ministero dell’Università e della Ricerca) and European Union (Next Generation EU) under the PRIN 2022 program (project code 2022CN4EBH) and the \textit{GENIO} project (CUP B69J23005770005) funded by MIMIT. 
We are grateful to our former student Martina Russo for her help in the early stages of this work.

%\section*{References}
%\IEEEtriggeratref{46}
\bibliographystyle{IEEEtran}
\bibliography{biblio.bib}

\end{document}